\documentclass[aps,prd,preprint,superscriptaddress,showpacs,showkeys]{revtex4}
\usepackage{amssymb,amsmath}
\usepackage{graphicx} 
\usepackage{dcolumn} 

\usepackage{epsfig}   
\usepackage{pstricks}
\usepackage{ifpdf}

\newcommand{\Tr}{ {\mathrm{Tr}\, }}

\begin{document}

\title{ On the Yang--Mills propagator at strong coupling}

\author{Y. Gabellini}
\affiliation{Institut de Physique de Nice, UMR 7010\\Universit\'{e} C\^ote d'Azur -- CNRS\\ 17 rue Julien Laupr\^ etre 06200 Nice, France}
\email[]{Yves.Gabellini@univ-cotedazur.fr, Thierry.Grandou@orange.fr}
\author{T. Grandou}
\affiliation{Institut de Physique de Nice, UMR 7010\\Universit\'{e} C\^ote d'Azur -- CNRS\\ 17 rue Julien Laupr\^ etre 06200 Nice, France}
\email[]{Thierry.Grandou@inphyni.cnrs.fr}
\author{R. Hofmann}
\affiliation{Institut f\"ur Theoretische Physik\\ 
Universit\"at Heidelberg\\ 
Philosophenweg 16 
69120 Heidelberg, Germany}
\email[]{r.hofmann@thphys.uni-heidelberg.de}

\date{\today}
\vspace{5cm}

\begin{abstract} 
About twelve years ago the use of standard functional manipulations was demonstrated to imply an unexpected property satisfied by the fermionic Green's functions of $QCD$. This non--perturbative phenomenon is dubbed \textit{Effective Locality}. In a much simpler way than in $QCD$, 
the most remarkable and intriguing aspects of Effective Locality have been presented in a recent letter in the Yang--Mills theory on Minkowski spacetime. While quickly recalled in the current paper, these results are used to calculate the problematic gluonic propagator in the Yang--Mills non--perturbative regime.\vspace{0.1cm}\\ 
{\sl This paper is dedicated to the memory of Professor Herbert M. Fried ( 1929 -- 2023 ) whose inspiring manner, impressive command of functional methods in quantum field theories, enthusiasm for a broad range of topics in Theoretical Physics, and warm friendship is missed greatly by the authors.}
\end{abstract}



\maketitle

\section{Introduction}

 A new property of QCD has been discovered fourteen years ago and named ``effective locality", or $EL$ for short \cite{EPJC}. In the very case of QCD, this property has often been phrased as follows :
 \par
 For any fermionic $2n$--point Green’s functions and related amplitudes, when the full sum of all interactions is completed, the resulting structure is that of a local contact--type interaction term. This local interaction is mediated by a tensorial field which is antisymmetric both in Lorentz and colour indices. Moreover, the sum just evoked happens to be gauge invariant \cite{tg}.
 \par
 The $EL$ property is non--perturbative and its derivation makes it clear that it is specifically due to non--abelian structure interactions : no possible track of it could ever be found in an abelian situation.
 \par
 The qualification of ``effective" refers to the fact that a complete integration of elementary gluonic degrees of freedom is carried out, while ``locality" refers to the unexpected contact--type resulting interaction. Does this mean that one should think of effective locality as a \emph{dual} form of the originally formulated $QCD$ theory ? The answer is certainly negative \cite{SW} in spite of preliminary results obtained thirty years ago in the euclidean Yang--Mills case and at first non trivial order of a semi--classical expansion \cite{RefF}. 
 \par
 Since its discovery, the property of effective locality has been explored both at theoretical and phenomenological levels. The $EL$ first outputs appear quite interesting and mostly in line with the expected features of the $QCD$ non--perturbative regime, but involve also some new predictions, more specific to the $EL$ property itself. For example, fermionic Green's functions have displayed an extra dependence on the trilinear Casimir operator of the $SU_c(3)$ algebra \cite{casimir}, which, however sub--leading, has elsewhere been recognised a hallmark of the non--abelian non--perturbative regime \cite{Dmitrasinovic}. These results have been summarised in a recent review article \cite{ygtg}.
 \par
 However, only the full $QCD$ case has been considered so far, as, in certain approximation schemes at least, the fermionic sector enjoys valuable simplifications;  and also because, elementary gluon fields having been integrated out, the effective locality form presents itself under the form of an effective quark model.
 \par
 Now, the full $QCD$ case obscures somewhat the most essential aspects of the $EL$ derivation which have to do with the gluonic sector of $QCD$. For example, the crucial issue of non--abelian gauge invariance, which would posit $EL$ as a sound and most promising property, can be read off the simpler Yang--Mills case. 
\par 
In a recent paper devoted to the Yang--Mills theory on Minkowski spacetime, the gluon Green's functions generating functional has been derived in full details in order to display the striking $EL$ property \cite{miely}. Concerning concrete calculations, however, the efficiency of the Yang--Mills gluon Green's functions generating functional obtained in this way can be questioned. 

The goal of the current paper is therefore to rely on the generating functional derived in \cite{miely} in order to get insights on the first and simpler Green's functions, the non--perturbative gluon propagator in the strong coupling regime.
\par 
 It may be worth recalling that the formal context of the whole matter is that of standard functional methods within Lagrangian quantum field theories \cite{Eik}.
\par\medskip
 The current paper is organised as follows. The next Section summarises as quickly as possible the derivation of \cite{miely}, while Section III takes advantage of it to estimate the gluonic propagator at strong coupling, with a result which clearly supports previous expectations \cite{Kondo}. Particular emphasis is put on the feasibility and rigour the $EL$ context offers to concrete calculations. Such an example is clearly offered by the subsection III.A in which the quite technical calculation of the one point Green's function is given in much details. A final Section IV concludes our paper.
 \par
 Of course, Yang--Mills theory has been the matter of an impressive amount of work and publications that the present paper does not pretend to list.

\section{Gluon Green's functions generating functional}

 In this section, a rapid summary of the essential steps at the origin of the effective locality form of the gluon Green's functions generating functional is given, whose full details can be found in \cite{miely}. 
 \par In Minkowski spacetime, the $YM$ Lagrangian density reads

 \begin{equation}\label{Lagrangien}
 {\mathcal{L}}_{Y\!M}(x)=-{1\over 4} F^a_{\mu\nu}(x)F^{a{\mu\nu}}(x)
 \end{equation}

\noindent where the $F^a_{\mu\nu}(x)$ are the customary non--abelian field strength tensors

  \begin{equation}
 F^a_{\mu\nu}(x)=\partial_{\mu}{A}_{\nu}^a(x) - \partial_{\nu}{A}_{\mu}^a(x) + gf^{abc}{A}_{\mu}^b(x){A}_{\nu}^c(x)
 \end{equation}

\noindent ${A}_{\mu}^a(x)$ being the gluon fields -- $a=(1,\cdots\,, 8)$  -- and where the $f^{abc}$ are the totally antisymmetric structure constants of the $SU_c(3)$--Lie algebra. 
An integration by parts yields

 \begin{equation}\label{identic}
\int  {\mathcal{L}}_{Y\!M}=-\frac{1}{4} \int{{F}^{2}} =  \frac{1}{2} \int{ \displaystyle{A}^a_{\mu}(x)\,\delta^{ab}\,\left(g^{\mu\nu}\,\partial^2 - \,\partial^{\mu}\partial^{\nu}\right){A}^b_{\nu}(x) } + \int{{\mathcal{L}}_{int}}[A]
\end{equation}

In a standard manner \cite{Eik}, this separation allows one to obtain the full gluon generating functional out of the free field one :

 \begin{eqnarray}\label{ZYM}
 Z_{YM}[j]&\equiv&\,_{\rm in}\!<\!0\,|\,T\,e\,^{\displaystyle\!-i\int\!\! d^4x\,j_{\mu}^a(x)\,A_{\mu}^a(x)}|\,0\!>_{\rm in}\nonumber\\ &=&N\, e\,^{\displaystyle{ i\int\!\!{\mathrm{d}}^4x\,{\cal L}_{\rm int}\bigl[{i}{\delta\over\delta j_{\rho}^c(x)}\bigr]}}\, e\,^{\displaystyle -{i\over2}\int\!\!\mathrm{d}^4x\,\mathrm{d}^4y\,\,j_{\mu}^a(x)\,D^{\mu\nu}_{ab}(x-y)\,j_{\nu}^b(y)}
 \end{eqnarray}

\noindent the constant $N$ being such that $Z_{YM}[0]=1$, and provided that the \emph{distribution} $D^{\mu\nu}_{ab}(x-y)$ exists. Now, this is the point, because $D^{\mu\nu}_{ab}$ is supposed to satisfy the equation

 \begin{equation}\label{inverse}
 \displaystyle \bigl[g^{\mu\rho}\partial^2_x - \partial_x^{\mu}\partial_x^{\rho}\bigr]\,D_{\rho}^{ab\, \nu}(x-y) = g^{\mu\nu}\delta^{ab}\,\delta^{(4)}(x-y)
 \end{equation}

As well known, the operator $\partial^2 -\partial\otimes\partial$ cannot be inverted and $D^{\mu\nu}_{ab}$ is accordingly not defined. In $QED$ as well as in perturbative $QCD$, solutions to this problem are well known, preserving Lorentz invariance and breaking momentarily the local gauge invariance of $ {\mathcal{L}}_{Y\!M}$ by means of a gauge--fixing term. In a second step, local gauge invariance is restored and checked through the Ward--Takahashi and Slavnov--Taylor identities satisfied by Green's functions, in the respective cases of $QED$ and $QCD$. All this is textbook material and defines the well controlled perturbative framework of these theories.

\par\bigskip
Now, the non--abelian structure of $ {\mathcal{L}}_{Y\!M}$ offers a possibility which has no abelian equivalent.  For example, adding and subtracting a term to (\ref{Lagrangien}) ${}^{\footnotemark[1]}$
\footnotetext[1]{This is but one example of derivation of the effective locality property, possibly the simplest. The property was originally derived by observing that contrarily to the case of  $QED$, in $QCD$, quantisation could be carried through by maintaining Lorentz and local gauge invariance at the same time \cite{EPJC}. }
 
 \begin{equation}\label{plusmoins}
 {\mathcal{L}}_{Y\!M}\ \longrightarrow\   {\mathcal{L}}_{Y\!M} + \frac{1}{2\zeta}(\partial\cdot \!{A})^2- \frac{1}{2\zeta}(\partial\cdot \!{A})^2
 \end{equation}

\noindent the original full gauge invariance of (\ref{Lagrangien}) is obviously preserved, while one of the two extra terms of (\ref{plusmoins}) can be used to turn the undefined expression of (\ref{inverse}) into the well defined covariant propagator ${{D}_{\mathrm{F}}^{(\zeta)}}$

 \begin{equation}\label{covprop}
 \left({{D}_{\mathrm{F}}^{(\zeta)}}^{-1}\right)_{\mu \nu}^{a b} = \delta^{a b} \, \left[ g_{\mu \nu} \, \partial^{2} - \left(1 -\frac{1}{\zeta} \right) \partial_{\mu} \partial_{\nu} \right]
\end{equation}
 
\par\medskip
Then, using (\ref{identic}) to re--express the density of interaction ${\mathcal{L}}_{int.}$, one arrives at

\begin{eqnarray}\label{ZYM1}
{Z}_{{YM}}[j] = {N} \, \left. e^{\displaystyle{-\frac{i}{4} \int{{F}^{2}} - \frac{i}{2}  \int{ \displaystyle{A}^a_{\mu}(x)\bigl(g^{\mu\nu}\,\partial^2 - \,(1 - \frac{1}{\zeta})\,\partial^{\mu}\partial^{\nu}\bigr){A}^a_{\nu}(x)}}} \right|_{\displaystyle A= \frac{1}{i} \, \frac{\delta}{\delta j} } \, e^{\displaystyle{-{\frac{i}{2} \int{j \cdot {D}_{\mathrm{F}}^{(\zeta)} \cdot j} }}}
\end{eqnarray}

A convenient rearrangement of (\ref{ZYM1}) can be obtained in a standard functional way \cite{Eik} :

\begin{eqnarray}\label{ZYM2}
{Z}_{{YM}}[j] &=& {N}\ e^{\displaystyle-{{\ \frac{i}{2} \int{j \cdot {D}_{\mathrm{F}}^{(\zeta)} \cdot j} }}}\nonumber\\ &\cdot& \left. e^{\ \displaystyle{{\mathfrak{D}^{(\zeta)} _{A}}}} \  e^{\displaystyle{{-\frac{i}{4} \int{{F}^{2}} - \frac{i}{2} \int{ A \cdot \left({ {D}_{\mathrm{F}}^{(\zeta)}}\right)^{-1} \cdot A} }}} \right|_{\displaystyle A = \int{{D}_{\mathrm{F}}^{(\zeta)}\cdot j} }
\end{eqnarray}

\noindent where the functional operator $\exp {\mathfrak{D}_{A}}$ reads

\begin{equation}\label{linkage}
\exp\mathfrak{D}^{(\zeta)} _{A} =  \exp\frac{i}{2} \int\mathrm{d}^4x\int\mathrm{d}^4y\ {\frac{\delta}{\delta A(x)} \cdot  {D}_{\mathrm{F}}^{(\zeta)}(x-y) \cdot \frac{\delta}{\delta A(y)} }
\end{equation}

\par\medskip
In order to proceed with doable exact calculations, one must introduce a ``linearization'' of the $\int F^2$ term which appears in the right hand side of (\ref{ZYM2}). This is achieved by using the famous Halpern's representation \cite{Halpern} (also known as the Hubbard--Stratonovich transformation in solid state physics \cite{HS}), which rests on the introduction of eight, antisymmetric in spacetime indices $\mu$ and $\nu$, tensor fields $ \chi^{a}_{\mu \nu}(x)$, $a=(1,\cdots\,, 8)$ :

\begin{equation}\label{Halpern}
e^{{\displaystyle{-\frac{i}{4} \int{{F}^{2}}}}} = N' \, \int{\mathrm{d}[\chi] \ e^{ {\displaystyle{\ \frac{i}{4} \int{\chi_{\mu \nu}^{a}\chi^{a\mu \nu} + \frac{i}{2} \int{ \chi^{a\mu \nu} {F}_{\mu \nu}^{a}} } } }}}
\end{equation}

\noindent allowing one to bring (\ref{ZYM2}) into the form

\begin{eqnarray}\label{ZYM3}
{Z}_{{YM}}[j] &=& {\cal N}\ e^{-\displaystyle{{\ \frac{i}{2} \int{j \cdot {D}_{\mathrm{F}}^{(\zeta)} \cdot j} }}}\int{\mathrm{d}[\chi] \ e^{ {\displaystyle{\ \frac{i}{4} \int{\chi_{\mu \nu}^{a}\chi^{a\mu \nu}}}}}}
\nonumber\\ &\cdot& \left. e^{\ \displaystyle{{\mathfrak{D}^{(\zeta)} _{A}}}} \  e^{\,\displaystyle{{\frac{i}{2} \int{\chi\cdot{F}} - \frac{i}{2} \int{ A \cdot \left({ {D}_{\mathrm{F}}^{(\zeta)}}\right)^{-1} \cdot A} }}} \right|_{\displaystyle A = \int{{D}_{\mathrm{F}}^{(\zeta)}\cdot j} }
\end{eqnarray}

\noindent where ${\cal N}=NN'$, and (\ref{ZYM3}) yields, as a final result, the gluon Green's functions generating functional of \cite{miely} :

\begin{eqnarray}\label{ZYM40}
&{Z}_{{YM}}[j] = {\cal N}\displaystyle\int{\mathrm{d}[\chi] \ e^{ {\displaystyle{\ \frac{i}{4} \int{\chi_{\mu \nu}^{a}\chi^{a\mu \nu}}}}}}\,\bigl[\det(gf\cdot\chi)\bigr]^{-\frac{1}{2}}
\nonumber\\ &\cdot  e^{\displaystyle -{i\over2}\int\!\!\mathrm{d}^4x\,\,\left(j_{\mu}^a(x)+\partial^\lambda\chi^a_{\lambda\mu}\right)(x)\,\bigl[(gf\cdot\chi)^{-1}\bigr]^{\mu\nu}_{ab}(x)\,\left(j_{\nu}^b(x)+\partial^\sigma\chi^b_{\sigma\nu}\right)(x)}
\end{eqnarray}

The normalisation condition ${Z}_{{YM}}[0]=1$ leads to :

 \begin{equation}\label{normal}
{\mathcal{N}}^{-1}=\int{\mathrm{d}[\chi] \ e^{ {\displaystyle{\ \frac{i}{4} \int{\chi_{\mu \nu}^{a}\chi^{a\mu \nu}}}}}}\!\!\bigl[\det(gf\cdot\chi)\bigr]^{-\frac{1}{2}}\,e^{\displaystyle -{i\over2}\int\!\!\mathrm{d}^4x\,\,\partial^\lambda\chi^a_{\lambda\mu}(x)\,\bigl[(gf\cdot\chi)^{-1}\bigr]^{\mu\nu}_{ab}(x)\,\partial^\sigma\chi^b_{\sigma\nu}(x)}
 \end{equation}

One may note that the form (\ref{ZYM40}) is not completely unknown. More than thirty years ago, it was observed in an attempt at deriving a form dual to (\ref{Lagrangien}) within an instanton based calculation \cite{RefF}; but most importantly, a final integration on the same $\chi$--field remained constrained by a gauge fixing condition with all of the ensuing well-known complications~\cite{Gribov}.

\section{Insights obtained out of the effective locality $Z_{YM}[j]$}

One may expect the effective locality $Z_{YM}[j]$ to provide interesting informations on the strong coupling non-perturbative regime of the $YM$ theory, and to lend itself to tractable calculations. This is what is investigated in the present section.

\subsection{Calculation of $\langle A_a^\mu(x)\rangle$}

This example will pave the route to the next section's considerations and set up the framework of effective locality calculations, while exhibiting the rigour of the calculation's developments offered by the $EL$ context.

\subsubsection {Presentation} 

From (\ref{ZYM40}) one obtains the ``one--point'' Green's function by means of one functional differentiation over ${Z}_{{YM}}$ :

\begin{eqnarray}\label{1pt}
&{\displaystyle\langle A_a^\mu(x)\rangle}=\displaystyle{ i}{\delta\over\delta j_{\mu}^a(x)}\,{Z}_{{YM}}[j]\Bigl|_{j=0} =  \displaystyle {\cal N}\,\int\!\!d[\chi]\,e\,^{\displaystyle  {i\over 4}\int\!\!d^4u\, ({\chi}_{\mu\nu}^a(u))^2}\,\bigl[\det (g\,f\!\cdot\!\chi)\bigr]^{-{1\over2}}\hfill\hfill \nonumber \\ &\displaystyle \cdot\,{1\over g}\,\partial_{\alpha}{\chi}_{\alpha\nu}^b(x)\,\,\bigl[(f\!\cdot\!\chi)^{-1}\bigr]^{ba}_{\nu\mu}(x)\,e\,^{\displaystyle{-{i\over 2}\int\!\!d^4u\,\partial_{\alpha}{\chi}_{\alpha\mu}^a(u)\,\bigl[( g\,f\!\cdot\!\chi)^{-1}\bigr]^{ab}_{\mu\nu}(u)\,\partial_{\beta}{\chi}_{\beta\nu}^b(u)}}\hfill
\end{eqnarray}

Calculating (\ref{1pt}) is a formidable task which could be undertaken in a future publication. For the current purpose, it is put to a more accessible form in the limit of strong coupling, $g\!>>\!1$. This is particularly consistent with the non--perturbative regime inherent to the $EL$ property and its (partial) duality aspect \cite{miely}. As will be proven elsewhere in the case of $QCD$, at first non trivial order at least, this property allows one to rely on an ultraviolet and infrared well behaved expansion in power of $g^{-1}$.
\par
At strong coupling, only the term of order $g^{-1}$ will be kept. The exponential, expanded in power of $g^{-1}$, is therefore taken to unity. At leading order,  $\langle A\rangle$ reads :

\begin{equation}\label{1pt'}
{\displaystyle\langle A_a^\mu(x)\rangle}= \displaystyle {\cal N}_1\int\!\!d[\chi]\,e\,^{\displaystyle  {i\over 4}\int\!\!d^4u\, ({\chi}_{\mu\nu}^a(u))^2}\,\bigl[\det (g\,f\!\cdot\!\chi)\bigr]^{-{1\over2}}\,\biggl( {1\over g}\biggr)\,\partial^{\alpha}{\chi}_{\alpha\lambda}^c(x)\,\,\bigl[(f\!\cdot\!\chi)^{-1}\bigr]_{ca}^{\lambda\mu}(x)
\end{equation}

At the same order of approximation, the normalisation is :

 \begin{equation}\label{normal'}
{\mathcal{N}}_1^{-1}=\int{\mathrm{d}[\chi] \ e^{ {\displaystyle{\ \frac{i}{4} \int{\chi_{\mu \nu}^{a}\chi^{a\mu \nu}}}}}}\!\!\bigl[\det(gf\cdot\chi)\bigr]^{-\frac{1}{2}}
 \end{equation}

 \subsubsection{A few formal steps: the matrix $\mathbb{M}$ and the measure image theorem}

   To proceed with concrete calculations, a series of previous analyses summarised in \cite{ygtg} have proven that it is convenient to take advantage of a(n analytically continued \cite{QCD6}) \emph{Random Matrix} integration \cite{Mehta} of (\ref{1pt'}) and (\ref{normal'}).  This is achieved by defining the hermitian matrix :

\begin{equation}\label{M}
\mathbb{M}(x)=i\displaystyle\sum_{a=1}^{8}\,\chi^a(x)\otimes T^a
\end{equation}

\medskip\noindent 
The $T^a$s are the eight Lie algebra generators of $SU_c(3)$ taken in the $8\times 8$ dimensional adjoint representation, where the matrix representation of these hermitian generators is given by the $SU_c(3)$--Lie algebra structure constants : 

\begin{equation}\label{T}
\bigl(T^a\bigr)^{bc}= -i\,f^{abc}
\end{equation}

Equations (\ref{M}) and (\ref{T}) lead to :

\begin{equation}\label{mat}
\bigl(\mathbb{M}\bigr)^{\mu\nu}_{ab} = \bigl(f\!\cdot\!\chi\bigr)^{\mu\nu}_{ab}
\end{equation}

One has now replaced the unmanageable $ \bigl(f\!\cdot\!\chi\bigr)^{\mu\nu}_{ab}$ term, and its $\chi$ integration, by an algebraic quantity that will turn out to be effective for computational purposes.

So, $\mathbb{M}$ is the sum of eight matrices, each of them being composed of $4\times4$ blocks of dimension $8\times8$. Its elements have the form $\bigl(\mathbb{M}\bigr)^{\mu\nu}_{ab}$ with both pairs of indices being antisymmetric. Nonetheless, $\mathbb{M}$ turns out to be a real traceless symmetric  matrice, of dimension $N\times N$, that is $32\times 32$, at $D=4$ spacetime dimensions and $N_c=3$ colours : $N=D(N_c^2-1)$. 
\par
Being real symmetric a matrix, $\mathbb{M}(x)$ can be diagonalised, $\mathbb{M}(x)={}^t\mathcal{O}(x)\,D(x)\,\mathcal{O}(x)$, where ${}^t\mathcal{O}(x)$ is the transpose of  $\mathcal{O}\in O_N(\mathbb{R})$, the orthogonal group, and where the matrix $D(x)$ is the \emph{real} diagonal matrix $D(x)=\mathrm{diag}\left(\xi_1,\xi_2,\dots,\xi_{N-1},\xi_N\right)$. Formally, one has also $\mathbb{M}^{-1}={}^t\mathcal{O}(x)\,D^{-1}(x)\,\mathcal{O}(x)$ with $D^{-1}(x)=\mathrm{diag}\left({\xi_1}^{-1},\,{\xi_2}^{-1},\,\dots,\,{\xi_{N-1}}^{-1},\,{\xi_N}^{-1}\right)$. With $\mathbb{M}(x)$ such as defined above, it is easy to check that for all point $x$ in $\mathcal{M}$, the eigenvalue $\xi=0$ belongs to the spectrum of $\mathbb{M}(x)$; this, however, will not induce difficulties in the subsequent integration processes \cite{ygtg}.

The key point here is that the property of effective locality allows one to rely on the \emph{measure image theorem} ${}^{\footnotemark[2]}$ \cite{Ted}\footnotetext[2]{When passing from an infinite dimensional functional space to a finite dimensional one, the {measure image theorem} is used and can be viewed as generalising the more customary notion of a \emph{Jacobian} between finite dimensional spaces \cite{ygtg}.}, in order to transform the initial functional measure of integration ${\mathrm{d}}[\chi]$ into the product of an integration on the spectrum of $\mathbb{M}$, ${\mathrm{Sp}}(\mathbb{M})$, times an integration on the orthogonal group $O_N(\mathbb{R})$; symbolically, 

\begin{equation}\label{mesprod} 
{\mathrm{d}} [\chi]\ \longrightarrow\ {\mathrm{d}}\,{\mathrm{Sp}}(\mathbb{M})\times {\mathrm{d}}\, O_N(\mathbb{R})
\end{equation}

In this way, the symbolic functional measure of integration $\mathrm{d}[\chi]$ with 

\begin{equation}\label{init-m}
\int{\mathrm{d}[\chi]} = \prod_{z\in\mathcal{M}} \prod_{a=1}^{N_c^2-1} \prod_{ 0= \mu<\nu}^3 \int {\mathrm{d}[\chi_{\mu \nu}^{a}(z)]}\,,
\end{equation}

\noindent where $\mathcal{M}$ is the spacetime manifold, is turned into a well defined measure of integration on the finite dimensional space of $\mathbb{M}$ matrices

\begin{eqnarray}\label{new-m}
{\rm{d}}\left(i\sum_{a=1}^{8}\,{\chi^a}(x)\otimes T^a\right)&\equiv& {\rm{d}}\mathbb{M}(x)= {\rm{d}}M_{11}\,{\rm{d}}M_{12} \cdots {\rm{d}}M_{NN} \nonumber \\ &=&\nonumber \left|\frac{ \partial(M_{11}, \cdots, M_{N\!N})}{\partial(\xi_1, \cdots, \xi_N, p_1, \cdots, p_{N(N-1)/2})}\right| \, {\rm{d}}\xi_1 \cdots {\rm{d}}\xi_N \, {\rm{d}}p_1 \cdots {\rm{d}}p_{N(N-1)/2} \\  &=& \prod_{i=1}^{N}\ {\rm{d}}\xi_i  \prod_{1\leq i<j}^N |\xi_i-\xi_j|^{\beta}\   {\rm{d}}p_1\  ..\ {\rm{d}}p_{N(N-1)/2}\, f({\mathbf{p}})
\end{eqnarray}

In this equation, $\beta = 1$, and  the very last factors of (\ref{new-m}) stand for a \emph{Haar measure} of integration on the orthogonal group $O_N(\mathbb{R})$. 

Note that in (\ref{new-m}), the value $\beta = 1$ could be bothering as it seems to compromise analyticity. This isn't so however, because it turns out that matrices $\mathbb{M}$ are also \emph{skew--symmetric}, possessing a spectrum formed of $N/2$ pairs of equal and opposite eigenvalues. For instance, they can be ordered in the following way :

\begin{equation}\label{A3}
{\mathrm{Sp}}\,\mathbb{M}=\bigl\lbrace\,(\xi_i,\xi_{N-i+1}=-\xi_i)\bigr\rbrace\ , \forall i=1,2,\dots, N/2\,.
\end{equation}

Taking this property into account, the apparently non--analytic \emph{Vandermonde determinant} can be rewritten as

\begin{eqnarray}\label{VdM}
\prod_{1\leq i<j}^N |\xi_i-\xi_j|^{\beta} =\biggl[\prod_{i=1}^{N/2}2|\xi_i|\biggr]^{\beta}\, \,\biggl[\prod_{1\leq i<j}^{N/2} (\xi^2_i-\xi^2_j)^2\,\biggr]^{\beta}
\end{eqnarray}

\noindent which, even at $\beta = 1$, will give rise to an analytic function of its eigenvalues because the first term in the right hand side of (\ref{VdM}), involving absolute values, is exactly compensated by the term $\bigl[\det(gf\cdot\chi)\bigr]^{-\frac{1}{2}}$. This applies to both (\ref{1pt'}) and (\ref{normal'}).

\subsubsection{Calculating $\langle A_a^\mu(x)\rangle$}

Going back to $\langle A_a^\mu(x)\rangle$, once one has replaced $(f\!\cdot\!\chi)_{ab}^{\mu\nu}(x)$ by $\bigl(\mathbb{M}\bigr)^{\mu\nu}_{ab}$  in (\ref{1pt'}), one needs then to deal with the $\partial^{\alpha}{\chi}_{\alpha\beta}^a(x)$ term, that has to be expressed in function of $\mathbb{M}$ too.

From definition (\ref{M}) and (\ref{T}), one can prove (and easily check in the simpler case of $SU(2)$) the following relation,

\begin{equation}\label{1}
\chi^c_{\alpha\lambda}(x)={1\over N_c}\,\Tr \bigl[\,\mathbb{M}(x)\, \left(\,\mathbb{I}_{\alpha\lambda}\otimes iT^c\,\right)\,\bigr]
\end{equation}

\noindent where the $4\times 4$ matrix $\mathbb{I}_{\alpha\lambda}$ has null matrix elements but for the element at row $\alpha$ and column $\lambda$, which has value $1$. In view of the antisymmetry of $\chi^c_{\alpha\lambda}$, one has necessarily $\alpha\neq \lambda$. The matrix $\mathbb{I}_{\alpha\lambda}$ is therefore non--symmetric, guarantying the non triviality of (\ref{1}), and has zeros on its diagonal :

\begin{equation}\label{proj}
{{}}(\mathbb{I}_{\alpha\lambda})^{\rho \rho}=0\,,\  \ \forall \rho=0,1,2,3\,.
\end{equation}

In order to provide calculations with a convenient covariant--like form, it is appropriate to pass from the $\mu,\nu=0,1,2,3$ spacetime indices and the $a,b=1,2,\dots,8$ colour indices, to indices of an enlarged linear space ${}^{\footnotemark[3]}$ \footnotetext[3]{Calculations are most easily carried out by endowing this enlarged linear space with the scalar product $g^{ij}=\delta^{ij}$, rather than the `more canonical' scalar product $g^{ij}=g^{\mu\nu}\otimes \delta^{ab}$ which would then substitute the pseudo--orthogonal group $O(8,24|\mathbb{R})$ to the simpler orthogonal $O_N(\mathbb{R})$ group with no definite gain but more involved calculations. See footnote ${\footnotemark[9]}$.}, of dimension $4\times 8=32$ ${}^{\footnotemark[4]}$ \footnotetext[4]{This operation was suggested already in Ref.\cite{Halpern}.}, on which matrices $\mathbb{M}(x)$ would operate as endomorphisms. One can define the \emph{covariant indices} $i$ and $r$ ${}^{\footnotemark[5]}$ \footnotetext[5]{Other mappings can be used and yield the same results.},

\begin{equation}\label{mapping}
i\,\equiv\,a+\mu n\,,\ \ \ \ \ r\,\equiv\,c+\lambda n\,,\ \ \ \ n\,\equiv N_c^2-1=8\,,\ \ \ \ \ i,r=1,2,\dots,32
\end{equation}

\noindent and these new indices relate to the previous ones through the one--to--one mapping, 

\begin{equation}\label{mappping}
(\mu,a)\rightarrow i=a+\mu n\,,\ \ \ \ \ \ i\rightarrow (\mu=\frac{i-i_{m^{(o)}n}}{n},\ a=i_{m^{(o)}n})\,,
\end{equation}

\noindent where  the notation $i_{m^{(o)}n}$ stands for \textit{$i$ modulo n}. The last terms of (\ref{1pt'}) can now be rewritten as

\begin{equation}\label{Amu2}
{1\over gN_c}\,\bigl[\mathbb{M}(x)^{-1}\bigr]_{ca}^{\lambda\mu}\,\partial^\alpha\,\Tr\, \left(\mathbb{M}(x)\,\mathbb{P}^{\,c}_{\alpha\lambda}\right)={1\over gN_c}\bigl[\mathbb{M}(x)^{-1}\bigr]^{ir}\partial^\alpha\,\Tr \left(\mathbb{M}(x)\,\mathbb{P}^{\,r}_{\alpha}\right)
\end{equation}

\par\medskip
\noindent where shorthands $\mathbb{P}^{\,c}_{\alpha\lambda}$ and $\mathbb{P}^{\,r}_{\alpha}$ are introduced for the $N\times N$ matrix $\mathbb{I}_{\alpha\lambda}\otimes iT^c$, which achieves the projection of $\mathbb{M}(x)$ on $\chi^c_{\alpha\lambda}(x)$ according to (\ref{1}). In view of (\ref{proj}), one has also

\begin{equation}\label{projector}
(\mathbb{P}^{\,r}_{\alpha})^{ii}=0\,,\  \ \forall i=1,2,.., 32\,.
\end{equation}

 Equation (\ref{Amu2}) right hand side can be written as :

\begin{equation}\label{Amu3}
{1\over gN_c}\,\bigl[\,({}^t{\mathcal{O}})^{iu}\,({\delta^{us}\over \xi_s})\,{\mathcal{O}}^{sr}\bigr]\,\partial^\alpha\,\bigl[\,({}^t{\mathcal{O}})^{hk}\,(\delta^{kl}\xi_k)\,\,{\mathcal{O}}^{lm}\,(\mathbb{P}^r_{\alpha})^{mh}\bigr]
\end{equation}

\noindent where the orthogonal matrix $\mathcal{O}$ and eigenvalues $\xi_is$ are taken at $x\in \mathcal{M}$ and all the indices but $i$ are summed over.
 
Carrying out the discrete summations on $u$ and $l$ in (\ref{Amu3}), one gets :

\begin{equation}\label{Amu4}
{1\over gN_c}\bigl[\mathbb{M}(x)^{-1}\bigr]^{ir}\partial^\alpha\,\Tr \left(\mathbb{M}(x)\,\mathbb{P}^{\,r}_{\alpha}\right)= {1\over gN_c}\,\bigl[\,\mathcal{O}^{si}\,{1\over \xi_s}\,\mathcal{O}^{sr}\bigr]\,\partial^\alpha\,\bigl[\,\xi_k\,\,\mathcal{O}^{kh}\mathcal{O}^{km}\,(\mathbb{P}^r_{\alpha})^{mh}\bigr]\,.
\end{equation}

\par
The $\partial^\alpha$ derivatives must be taken before integration on $O_N(\mathbb{R})$ and $\mathrm{Sp}(\mathbb{M})$, and bear on $\mathcal{O}^{kh}\mathcal{O}^{km}(x)$ and on $\xi_k(x)$. 

\par
In either cases, in order to proceed, it will be essential to work in a basis where the orthogonal matrix is block--diagonal, the blocks being $2\times 2$ matrices of rotation in $2$--dimensional subspaces mutually orthogonal to each others. The orthonormal basis of the $\mathbb{M}(x)$ matrix eigenvectors can be used, where $\mathbb{M}(x)$ can be expanded as $\sum_{k=1}^N\, \xi_k\,u_k{}^tu_k$, and the basis vectors re--ordered in such a way that in the new ordered proper basis one has $\mathbb{M}(x)=diag\,\left(\,(\xi_1,\xi_2= -\xi_1), (\xi_3, \xi_4=-\xi_3), \dots, (\xi_{N-1}, \xi_N=-\xi_{N-1})\,\right)$. A linear algebra theorem \cite{Rodrigues} then asserts that $N/2$ angles exist, $\Theta_i=1,3,5,\dots N-1$ such that each of them parametrises a rotation or an inversion, in a given 2--dimensional subspace. 
\par
For example in the subspace spanned by the two eigenvectors $u_{2k+1}, u_{2k+2}$ corresponding to the eigenvalues $\xi_{2k+1}$ and $\xi_{2k+2}=-\xi_{2k+1}$, one will have the $2\times 2$ rotation $R(\Theta_{2k+1})$, or the inversion $I(\Theta_{2k+1})$, as the $(k+1)$--th block diagonal matrix representing, in this basis, the orthogonal matrix $\mathcal{O}$, with $0\leq k\leq N/2-1$.

\begin{equation}\begin{matrix}
0&\dots&0&\cos\Theta_{2k+1}  -\sin\Theta_{2k+1} & 0&\dots&0\\
0&\dots&0&\sin\Theta_{2k+1}  \,\,\,\cos\Theta_{2k+1} & 0&\dots&0\\
0&\dots&0 &\!\!\!\!\!\!\!\!\!\!\!\!\!\!\!\!\!\!\!\!\!\!\!\!\!\!\!\!\!\!0&\!\!\!\!\!\!\!\!\!\!\!\!\!\!\!\!\!\!\!\!\!\!\!\!\!\!\!\!\!\!0&\ddots\\
0&\dots&0
\end{matrix}\end{equation}

Each row and each column of the orthogonal matrix $\mathcal{O}$, contains only 2 non--zero coefficients that are either $\cos\Theta_{2k+1}$ and $-\sin\Theta_{2k+1}$ or $\cos\Theta_{2k+1}$ and $\sin\Theta_{2k+1}$ in the both cases of a rotation and an inversion whose matrix reads as 
 
$$I(\Theta_{2k+1})= \begin{pmatrix} 
\cos\Theta_{2k+1}\  \sin\Theta_{2k+1}\\
\sin\Theta_{2k+1} \,-\cos\Theta_{2k+1}\\
\end{pmatrix}$$

\par\medskip
Getting back to (\ref{Amu4}), a  first contribution of $\partial^\alpha$ acting on $\xi_k$ only is thus given by the expression

\begin{equation}\label{Amu5}
{1\over gN_c}\,\Bigl[{\partial^\alpha \xi_k\over \xi_i}\Bigr]\,\,\bigl[{O}^{ii}\,\,{O}^{ir}\,\,{O}^{kh}\,{O}^{km}\,\bigr]\,(\mathbb{P}^r_{\alpha})^{mh}\,,
\end{equation}

\par\medskip
Integration on $O_N(\mathbb{R})$ can be performed -- see \cite{Zhang} for instance --  yielding :

\begin{equation}\label{dalpha}
[{\partial^\alpha \ln\xi_i}] \,\bigl[\delta^{ir}\delta^{hm}+\delta^{ih}\delta^{rm}+\delta^{im}\delta^{rh}\bigr]\,(\mathbb{P}^r_{\alpha})^{mh} =[{\partial^\alpha \ln\xi_i}]\,\sum_r\,\bigl[(\mathbb{P}^r_{\alpha})^{ri}+(\mathbb{P}^r_{\alpha})^{ir}\bigr],
\end{equation}

\noindent where (\ref{projector}) is used for the first couple of $\delta$--constraints. For the two remaining terms in the right hand side of (\ref{dalpha}), one has :

\begin{equation}\label{PO}
(\mathbb{P}^r_{\alpha})^{ri}=(\mathbb{P}^r_{\alpha})^{ir}=0
\end{equation}

This can be proven by inspection. The index $i$ is fixed and choices of indices $\alpha$ and $\lambda$ can be made among $(16-4=12)/2=6$ possibilities. A given choice of $\alpha$ and $\lambda$ decides of the location of the $8\times 8$ matrix $T^c$, among the sixteen $8\times 8$ blocks composing the full projector $\mathbb{P}^c_{\alpha\lambda}$, the diagonal being excluded ${}^{\footnotemark[6]}$ \footnotetext[6]{One may observe that over the $1024$ matrix elements of the projector $\mathbb{P}^c_{\alpha\lambda}$, only $12$ of them, at most, are non zero.}. 

\par
The matrix elements of $(\mathbb{P}^r_{\alpha})=\mathbb{P}^c_{\alpha\lambda}$ with entries $ir$ or $ri$ are either zero systematically (see footnote [6]), or non zero \textit{a priori}. In this latter case, however, at entries $ir$ or $ri$, the matrix elements of $\mathbb{P}^c_{\alpha\lambda}$ coincide with elements of the matrix of $T^c$, taken at two identical entries and this is zero by antisymmetry of the generators $T^c$. This is seen as follows. One has, for all index $i$, for all $c=1,2,\dots,8,$ and all $r=1,2,\dots,N$, 

\begin{equation}\label{O}
r=c+\lambda n\Longleftrightarrow r=c\, \,m{}^{(o)}n\Longrightarrow(\mathbb{P}^c_{\alpha\lambda})^{ir}=(T^c)^{ic}=0\,,
\end{equation}

\noindent In the specific case of $\alpha=2$ and $\lambda=1$, for instance, $(\mathbb{P}^c_{\alpha\lambda})^{ir}$ is identically $0$ by construction, in the most trivial situation where $i>8$. This is why in the last term of (\ref{O}), the ``covariant'' index $i$ is preserved, in a somewhat \emph{hybrid} way, instead of $i\,m{}^{(o)}n$ of (\ref{mapping}). This applies identically to the reverse entries $ri$. 
\par\medskip
The other contribution to (\ref{Amu4}) reads,

\begin{equation}\label{other}
{1\over gN_c}\,{O}^{si}\,{O}^{sr}\ \partial^\alpha {O}^{sh}\,{O}^{sm}\,(\mathbb{P}^r_{\alpha})^{mh}
\end{equation}

This reduction of (\ref{Amu4}) to (\ref{other}) is due to the fact that if $k\neq s$, then the odd character of the $\xi_k$--integration yields a vanishing result~${}^{\footnotemark[7]}$ \footnotetext[7]{To see this one must transform the integration variables according to $\forall i,\,\xi_i\rightarrow -\xi_i$, and not $\xi_k$ solely. In this way, the overall integration on $\mathrm{Sp}(\mathbb{M})$ goes into minus itself, while preserving the whole set of the skew--symmetric constraints $\prod_{i=1}^{N/2}\delta(\xi_i+\xi_{N-i+1})$ introduced after (\ref{2011}). }. 

\par\bigskip
Now, the index $i$ being fixed, the sum on $s,h,m$ in (\ref{other}) is very restricted : for an even value of $i$ one gets in effect,

\begin{equation}\label{une}
{O}^{i-1,i}{O}^{i-1,i-1}\partial^\alpha\,{O}^{i-1,i}{O}^{i-1,i-1}\,(\mathbb{P}^{i-1}_{\alpha})^{i-1,i}\,+\,{O}^{ii}{O}^{i,i-1}\partial^\alpha\,{O}^{ii}{O}^{i,i-1}\,(\mathbb{P}^{i-1}_{\alpha})^{i,i-1}=0
\end{equation}

\noindent and a similar expression for an odd value of index $i$,

\begin{equation}\label{deux}
{O}^{ii}{O}^{i,i+1}\partial^\alpha\,{O}^{ii}{O}^{i,i+1}\,(\mathbb{P}^{i+1}_{\alpha})^{i,i+1}\,+\,{O}^{i+1,i}{O}^{i+1,i+1}\partial^\alpha\,{O}^{i+1,i}{O}^{i+1,i+1}\,(\mathbb{P}^{i+1}_{\alpha})^{i,i+11}=0\,.
\end{equation}

Both expressions are cancelling because of (\ref{O}), but would also yield a vanishing result upon integration over the angles $\Theta_{2k+1}$.
\par \noindent
At the end, because of (\ref{dalpha}), (\ref{PO}), (\ref{une}) and (\ref{deux}), one finds :

\begin{equation}\label{Amu0}
\langle A_a^\mu(x)\rangle=0\,.
\end{equation}

\subsection{The Yang--Mills propagator at strong coupling}

\subsubsection {Presentation}

The gluon 2--point Green's function is obtained from (\ref{ZYM40}) by means of two functional differentiations,

\begin{equation}
\displaystyle \langle T(A_{\mu}^a(x)A_{\nu}^b(y))\rangle\,= \, i{\delta\over\delta j_{\mu}^a(x)} i{\delta\over\delta j_{\nu}^b(y)}\,{\cal Z}_{YM}[j]\Bigl|_{j=0}
\end{equation}

\noindent and reads accordingly,

\begin{eqnarray}\label{2pts}
&\langle T(A_{\mu}^a(x)A_{\nu}^b(y))\rangle =\displaystyle  {\mathcal{N}}\int{\mathrm{d}[\chi] \ e^{ {\displaystyle{\ \frac{i}{4} \int{\chi_{\mu \nu}^{a}\chi^{a\mu \nu}}}}}}\,\bigl[\det(gf\cdot\chi)\bigr]^{-\frac{1}{2}}
\nonumber\\ &\cdot  \Bigl[ {i}\,\bigl[( g\,f\!\cdot\!\chi)^{-1}\bigr]^{ab}_{\mu\nu}(x)\,\delta^{(4)}(x-y) +\,\partial^{\alpha}{\chi}_{\alpha\lambda}^c(x)\,\bigl[( g\,f\!\cdot\!\chi)^{-1}\bigr]^{ca}_{\lambda\mu}(x)\,\,\partial^{\beta}{\chi}_{\beta\rho}^d(y)\,\bigl[(g\,f\!\cdot\!\chi)^{-1}\bigr]^{db}_{\rho\nu}(y)\,\Bigr] \nonumber \\ &\cdot\, e^{\displaystyle -{i\over2}\int\!\!\mathrm{d}^4x\,\,\,\partial^\lambda\chi^a_{\lambda\mu}(x)\,\bigl[(gf\cdot\chi)^{-1}\bigr]^{\mu\nu}_{ab}(x)\,\partial^\sigma\chi^b_{\sigma\nu}(x)} 
\end{eqnarray}

At strong coupling, the first term in the squared brackets is of order $g^{-1}$, while the second term is of order $g^{-2}$ and is thus a sub--leading correction. The leading order gluon propagator is therefore given by the simpler expression :

\begin{equation}\label{2pts2}
\langle T(A_{\mu}^a(x)A_{\nu}^b(y))\rangle = {\mathcal{N}_1}\,{ i\over g}\,\delta^{(4)}(x-y)\!\int\!\!{\mathrm{d}[\chi] \ e^{ {\displaystyle{\ \frac{i}{4} \!\int{\chi_{\mu \nu}^{a}\chi^{a\mu \nu}}}}}}\bigl[\det(gf\cdot\chi)\bigr]^{-\frac{1}{2}}\bigl[(f\!\cdot\!\chi)^{-1}\bigr]^{ab}_{\mu\nu}(x)
\end{equation}

 \noindent where the exponential, expanded in power of $g^{-1}$ is therefore taken to unity and the normalisation factor ${\mathcal{N}_1}$ given by  (\ref{normal'}).
\par
That is, if the right hand sides of both (\ref{2pts2}) and (\ref{normal'}) are well defined, then the leading order contribution (\ref{2pts2}) to the $YM$ gluon propagator at strong coupling would indicate an absence of propagation and a possible gluon condensate : as recalled in our concluding remarks, the importance of this conclusion has been analysed and stressed in \cite{Kondo} in regards of the crucial non--perturbative properties of \emph{mass gap} \cite{Douglas} and \emph{quark confinement}.

\subsubsection {Using the matrix $\mathbb{M}$}

Obviously, the $\chi$ integral of (\ref{2pts2}) differs from the one of (\ref{1pt'}), the divergence term $\partial^{\alpha}{\chi}_{\alpha\lambda}^c(x)$ being absent in (\ref{2pts2}). This will lead to very different kinds of computations between $\langle A_a^\mu(x)\rangle$ and $\langle T(A_{\mu}^a(x)A_{\nu}^b(y))\rangle$.

For (\ref{2pts2}) and (\ref{normal'}), integration on the functional space of $\chi^a_{\mu\nu}$ field conurations is now translated into an integration on $\mathbb{M}$, as in the previous section, and using formulas (\ref{mat}) to (\ref{VdM}), one obtains :

\begin{eqnarray}\label{201}
\langle T(A_{\mu}^a(x)A_{\nu}^b(y))\rangle = {\mathcal{N}_1}\,{ i\over g}\delta^{(4)}(x-y)\frac{1}{g^{N/2}}\prod_{i=1}^{N}\ \int^{+\infty}_{-\infty} \frac{{\rm{d}}\xi_i}{\sqrt{\xi_i}} \,\prod_{i=1}^{N/2}\delta\,(\xi_i+\xi_{N-i+1})\nonumber\\ \cdot\,\prod_{j=1}^{N/2}2|\xi_j|\, \prod_{1\leq k<l}^{N/2} (\xi^2_k-\xi^2_l)^2\, e^{ {\displaystyle{\ \frac{i}{8N_c}\sum_1^N\Delta\xi_i^2 }}}\, \int_{O_N(\mathbb{R})}\mathrm{d}O\,{\left({\mathbb{M}}^{-1}(x)\right)}^{ab}_{\mu\nu}\,,
\end{eqnarray}

\noindent and in the same way, for (\ref{normal'}),

\begin{eqnarray}\label{2011}
{\mathcal{N}}^{-1}_1=\frac{1}{g^{N/2}}\prod_{i=1}^{N}\ \int^{+\infty}_{-\infty} \frac{{\rm{d}}\xi_i}{\sqrt{\xi_i}} \,\prod_{i=1}^{N/2}\delta\,(\xi_i+\xi_{N-i+1})\nonumber\\ \cdot\,\prod_{j=1}^{N/2}2|\xi_j|\, \prod_{1\leq k<l}^{N/2} (\xi^2_k-\xi^2_l)^2\, e^{ {\displaystyle{\ \frac{i}{8N_c}\sum_1^N\Delta\xi_i^2 }}}\,\int_{O_N(\mathbb{R})}\mathrm{d}O\,.
\end{eqnarray}

Because the eigenvalues $\xi_is$ are dimensionful quantities, extra dimensions are brought in by the skew--symmetric constraints, the $\delta\,(\xi_i+\xi_{N-i+1})$, which cancel out in the ratio of (\ref{201}) to (\ref{2011}) and are accordingly omitted. In the argument of the exponential term, a parameter $\Delta$ with dimension of a (length)$\,{}^4$ must be introduced to keep the argument dimensionless (see comments (ii) and (iii) below).

\par
As done in the previous section, the original matrix element ${\left(\mathbb{M}^{-1}\right)}^{ab}_{\mu\nu}$ will be replaced by the new matrix element $({\mathbb{M}^{-1}})^{ij}$ with new indices $i=a+n\mu$ and $j=b+n\nu$  (see (\ref{mapping})).
The right hand side of (\ref{201}), can therefore be written as,

\begin{equation}\label{202}
{\mathcal{N}_1}\,{ i\over g}\delta^{(4)}(x-y)\,\frac{2^{N/2}}{g^{N/2}}      \,\prod_{i=1}^{N/2}\ \int^{+\infty}_{-\infty} {{\rm{d}}\xi_i} \prod_{1\leq i<j}^{N/2} (\xi^2_i-\xi^2_j)^2\, e^{ {\displaystyle{\ \frac{i}{4N_c}\sum_1^{N/2}\Delta\,\xi_i^2 }}} \!\int_{O_N(\mathbb{R})}\mathrm{d}O\,( {}^tO)^{ik}\, D^{-1}_{kl}\, O^{lj}\,.
\end{equation}

The last integration on $O_N(\mathbb{R})$ is straightforward and yields \cite{casimir},

\begin{equation}\label{ON} 
\int_{O_N(\mathbb{R})}\!\!\mathrm{d}O\,\sum_{k=1}^N( {}^tO)^{ik}\, D^{-1}_{kl}\, O^{lj}=\int_{O_N(\mathbb{R})}\!\!\mathrm{d}O\,\sum_{k=1}^N( {}^tO)^{ik}\, {\delta_{kl}\over \xi_k+i\varepsilon}\, O^{lj}=-i\pi\,\delta^{ij}\,\mathrm{vol}\,(O_N(\mathbb{R}))\,\sum_{k=1}^{{N/2}}\,\delta(\xi_k)
\end{equation}

\par\medskip
\noindent where in the right hand side the full volume of the orthogonal group $O_N(\mathbb{R})$ appears \cite{Zhang},

$$\mathrm{vol}\,(O_N(\mathbb{R}))=\frac{2^N\pi^{N(N+1)/4}}{\prod_1^N\Gamma({k/2})}$$

\noindent and where the poles at $\xi_k=0$ are displaced by an infinitesimal amount of $+i\varepsilon$ and the usual identity has been used, ${1}/{(x+i\varepsilon)}= v.p.({1}/{x})-i\pi\delta(x)$, the first term of which being the \emph{Cauchy's principal value} distribution ${}^{\footnotemark[8]}$ \footnotetext[8]{On the real field $\mathbb{R}$, integration on $\xi_k$s can only be defined through a principal value distribution, and the result is pure imaginary. This complies with the complex field $\mathbb{C}$--analysis where one can prove the existence of the $\xi_k$ integration in (\ref{202}) by analytical continuation of the \emph{Meijer}'s special functions \cite{QCD6}.}. 

\par
The eigenvalues $\xi_i$ have dimension (mass)${}^2$, {\textit{i.e.}}, $\xi_i=\Lambda^2\xi'_i$, with $\Lambda$ some mass scale. Keeping the same symbol $\xi_i$, the right hand side of (\ref{202}) can eventually be written in terms of dimensionless integration variables,

\begin{eqnarray}\label{203}
&\langle T(A_{\mu}^a(x)A_{\nu}^b(y))\rangle =\displaystyle{\mathcal{N}}_1\,{\pi\over g}\delta^{(4)}(x-y)\,\delta^{ij}\frac{N}{2}\Lambda^{-2} \,\mathrm{vol}\,(O_N(\mathbb{R}))\,\frac{2^{{N}/{2}}}{g^{{N}/{2}}}\,\,\Lambda^N\Lambda^{{N(N-1)(N-2)/2}}\nonumber \\ & \cdot\displaystyle\int^{+\infty}_{-\infty} \!{{\rm{d}}\xi_1}\, \delta(\xi_1)\int^{+\infty}_{-\infty} {{\rm{d}}\xi_2}\dots\int^{+\infty}_{-\infty} {{\rm{d}}\xi_{N\over 2}}\prod_{1\leq k<l}^{{N}/{2}} (\xi^2_k-\xi^2_l)^2\, e^{ {\displaystyle{\ \frac{i}{4N_c}\Delta\Lambda^4\sum_1^{{N}/{2}}\xi_i^2 }}}
\end{eqnarray}

\par\noindent and the normalisation can be calculated in the same way,

\begin{eqnarray}\label{N}
&{\mathcal{N}}^{-1}_1=\displaystyle\mathrm{vol}\,(O_N(\mathbb{R}))\,.\,\frac{2^{{N}/{2}}}{g^{{N}/{2}}}\, .\,\Lambda^N\Lambda^{{N(N-1)(N-2)/2}}  \nonumber \\ &\displaystyle\cdot\int^{+\infty}_{-\infty} \!{{\rm{d}}\xi_1}\int^{+\infty}_{-\infty}\! {{\rm{d}}\xi_2}\dots\int^{+\infty}_{-\infty} \!{{\rm{d}}\xi_{N\over 2}}\prod_{1\leq k<l}^{{N}/{2}} (\xi^2_k-\xi^2_l)^2\, e^{ {\displaystyle{\ \frac{i}{4N_c}\Delta\Lambda^4\sum_1^{{N}/{2}}\xi_i^2 }}}
\end{eqnarray}

\par\medskip
At the $g^{-1}$ leading order, the result for the Yang--Mills propagator at strong coupling can therefore be put into the following form,

\begin{equation}\label{204}
\langle T(A_{\mu}^a(x)A_{\nu}^b(y))\rangle =\ \frac{\pi N}{2\,g}\ \frac{I_1(\Delta\Lambda^4)}{I_2(\Delta\Lambda^4)}\ \frac{1}{\Lambda^2}\ \delta^{ab}\,{g_{\mu\nu}}\,\delta^{(4)}(x-y)\ +\,{\mathcal{O}}({1\over g^2})\,,
\end{equation}

\par\medskip
\noindent where the functions $I_1$ and $I_2$ of $\Delta\Lambda^4$ are the integrals which appear in the right hand sides of the expressions (\ref{203}) and (\ref{N}) respectively, and where the mapping (\ref{mappping}) has been used to recover the original spacetime and internal colour indices ${}^{\footnotemark[9]}$ \footnotetext[9]{In full rigour one should rather write $\delta^{ab}\,{\delta_{\mu\nu}}$ in view of our choice of the scalar product $g^{ij}\equiv\delta^{ij}$ (see footnote [3]). That is, $g_{\mu\nu}$ is ``put by hand'' in order to comply with the expected covariance of the gluon propagator. This is the price to be paid in order to deal with the simpler orthogonal group $O_N(\mathbb{R})$.}.
\par\medskip\noindent
Some preliminary remarks are in order. 
\par
(i)  Result (\ref{204}) is clearly non perturbative, as it should. Not only because it is on the order of $g^{-1}$, but as advertised before, because it indicates also an absence of propagation, in sharp contradistinction to the perturbative regime where gluon degrees of freedom propagate.
\par\medskip
(ii)  In (\ref{204}), two parameters show up: a mass, $\Lambda$, and a correlated dimensionless parameter $\Delta\Lambda^4$. In these derivations, nothing determines what $\Lambda$ should precisely be. However, that such a mass term should be present is a clear output of the current analysis and this is a non trivial point for a theory which is massless from the start \cite{Douglas}. The only mass scale one can think of in a quantum Yang--Mills theory is that of the \emph{asymptotic freedom} $\Lambda_{YM}$ parameter say, which is a renormalisation group invariant. There shouldn't be any inconsistency in assuming the relation $\Lambda\simeq\Lambda_{YM}$, as $\Lambda_{YM}$ is the scale below which non--perturbative effects are expected to come into play.
\par\medskip
(iii) The second parameter, $\Delta\Lambda^4$, is a pure numerical factor which involves some volume element $\Delta$ inherent to the definition of the initial measure (\ref {init-m}) on the spacetime manifold $\mathcal{M}$. In a canonical way \cite{Zee}, in effect, the measure (\ref {init-m}) definition assumes a decomposition of $\mathcal{M}$ into a collection of elementary cells of infinitesimal volume $\delta$, centered around each point of $\mathcal{M}$. The extension $\delta$, a \emph{meshing parameter}, could therefore appear to be the infinitesimal element met in mathematics, which is a formal \emph{indefinite} quantity, that is, deprived of any definite measure. In physical theories though, things must often be re--interpreted. In the current case, the formal {indefinite} elementary volume extension $\delta$, must be turned into a physical elementary volume extension $\Delta$, relative to the effective locality distance scale \cite{tgpt}: given 2 points $x$ and $y$ in $\mathcal{M}$, in effect, fields' correlations are sensitive to non--perturbative effects beyond a distance $L\geq \Lambda^{-1}_{YM}$ only. At this point one may observe that such a situation would normally preclude an interpretation of the effective locality form as a form dual to the original $YM$ case, at least in the most canonical definition of duality.
 
\subsubsection{Computations: Definiteness and calculability }
To begin with, a simple rescaling of the integration variables is enough to determine the dependence of the ratio $I_1/I_2$ on the parameter $\Delta\Lambda^4$. One obtains,

\begin{equation}\label{205}
\langle T(A_{\mu}^a(x)A_{\nu}^b(y))\rangle =\frac{\pi N}{2\,g}\ \frac{C_1}{C_2}\ \sqrt{\frac{\Delta}{4N_c}}\ \delta^{ab}\,{g_{\mu\nu}}\,{\delta^{(4)}(x-y)}\ +\,{\mathcal{O}}({1\over g^2})
\end{equation} 

\noindent where,

\begin{equation}\label{C1}
C_1= \int^{+\infty}_{-\infty} \!{{\rm{d}}\xi_1}\,\delta(\xi_1)\int^{+\infty}_{-\infty}\! {{\rm{d}}\xi_2}\dots\int^{+\infty}_{-\infty} \!{{\rm{d}}\xi_{N/2}}\prod_{1\leq k<l}^{{N}/{2}} (\xi^2_k-\xi^2_l)^2\, e^{ {\displaystyle{\ {i}\sum_1^{{N}/{2}}\xi_i^2 }}}
\end{equation}

\noindent and,

\begin{equation}\label{C2}
C_2= \int^{+\infty}_{-\infty} \!{{\rm{d}}\xi_1}\int^{+\infty}_{-\infty}\! {{\rm{d}}\xi_2}\dots\int^{+\infty}_{-\infty} \!{{\rm{d}}\xi_{N/2}}\prod_{1\leq k<l}^{{N}/{2}} (\xi^2_k-\xi^2_l)^2\, e^{ {\displaystyle{\ {i}\sum_1^{{N}/{2}}\xi_i^2 }}}
\end{equation}

Note that in (\ref{205}) the mass parameter $\Lambda$  has disappeared to the exclusive benefit of the meshing parameter $\Delta$. 
\par
This striking result indicates that if the meshing parameter is sent to zero, then, the non--perturbative effective locality contributions to the gluon 2--point function at strong coupling simply vanish. This is of course consistent with the $YM$ asymptotic freedom property  relevant to the short distance limit which is that of $\Delta\rightarrow 0$.
\par
Hence, as compared to the previous expression (\ref{204}), (\ref{205}) displays in a more cogent manner that the meshing parameter cannot be the indefinite $\delta$ parameter if (\ref{205}) is to make sense. And rightly so, as discussed above in remark (iii), the meshing parameter relates to the $YM$ scale $\Lambda_{YM}$ through the simple relation $\Delta\Lambda_{YM}^4\geq 1$.
\par
If (\ref{205}) is to make physical sense now, the ratio $C_1/C_2$ must be defined and computable. 
 \par\noindent
 For immediate purpose Cauchy's theorem is used on contour of Fig.1 in order to establish the analytical continuation of the $\xi_k$ to the integration variables $\Theta_k$ : $\xi_k \rightarrow\sqrt{i}\,\Theta_k$, with $\Theta_k\in \!\mathbb{R}$, \cite{QCD6}. One obtains,

\begin{figure}[!htb]
 \includegraphics[width=0.6\linewidth]{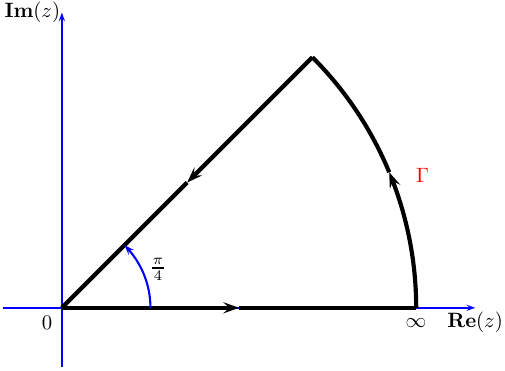}
 \caption{The integration contour displaying the analyticity domain of the integrals of (\ref{C1}) and (\ref{C2}) }
\label{Fig.1}
\end{figure} 

\begin{equation}\label{I1}
C_1={\sqrt{i}}^{{N}/{2}-1} \int^{+\infty}_{-\infty} \!{{\rm{d}}\Theta_1}\, \delta(\Theta_1)\int^{+\infty}_{-\infty}\! {{\rm{d}}\Theta_2}\dots\int^{+\infty}_{-\infty} \!{{\rm{d}}\Theta_{N/2}}\prod_{1\leq k<l}^{{N}/{2}} (\Theta^2_k-\Theta^2_l)^2\, e^{ {\displaystyle{\ -\sum_1^{{N}/{2}}\Theta_i^2 }}}\,.
\end{equation}

\noindent and likewise for $C_2$,

\begin{equation}\label{I2}
C_2={\sqrt{i}}^{{N}/{2}}\, \int^{+\infty}_{-\infty} \!{{\rm{d}}\Theta_1}\int^{+\infty}_{-\infty}\! {{\rm{d}}\Theta_2}\dots\int^{+\infty}_{-\infty} \!{{\rm{d}}\Theta_{N/2}}\prod_{1\leq k<l}^{{N}/{2}} (\Theta^2_k-\Theta^2_l)^2\, e^{ {\displaystyle{\ -\sum_1^{{N}/{2}}\Theta_i^2 }}}
\end{equation}

The first output of this analytical continuation is to make clear that $C_1$ and $C_2$ are non--vanishing numbers. The ratio $C_1/C_2$ is therefore a well defined number, 

\begin{equation}
{C_1\over C_2}=\frac{1}{\sqrt{i}}\,|{C_1\over C_2}|\,.
\end{equation} 

\medskip
However, the difficulty in calculating $C_1$ and $C_2$ lies in the Vandermonde determinant $\prod_{1\leq k<l}^{{N}/{2}} (\Theta^2_k-\Theta^2_l)^2$ which entails as much as $2^{{N(N-2)/8}}$ monomials, alternate in signs, that is $2^{120}$ terms. 
\par
Each monomial of the Vandermonde expansion might be integrated out exactly over the whole ${\mathrm{Sp}}(\mathbb{M})$, relying on integration formula,

 $$\int_{-\infty}^{+\infty} \mathrm{d}\Theta\ \Theta^{2p}\, e^{-a\,\Theta^2} =\frac{(2p-1)!!}{(2a)^p}\,\sqrt{\frac{\pi}{a}}$$

\medskip
Though not encountering any difficulty of principle.
 it remains that the control of such a huge number of terms, alternate in signs, is a formidable task, most likely out of reach practically.

\subsubsection{Connecting to the Random Matrix formalism}

 On the other hand, the analytical continuation $\xi_k \rightarrow\sqrt{i}\,\Theta_k$ just introduced makes it possible to rely on the powerful \emph{Random Matrix Theory} and the \emph{Wigner's semicircle law} in order to bound the constants $C_1$ and $C_2$ within definite intervals.  
\par
In order to see this, some Random Matrix Theory definitions must be introduced \cite{Mehta}, 

\begin{equation}\label{M1}
P_{N\beta}(\Theta_1,\dots,\Theta_N)\equiv \ C_{N\beta}\prod_{ i<j}^N|\Theta_l-\Theta_j|^\beta \,e^{{-}{\beta/2}\sum_1^N\Theta_i^2}\,,
\end{equation}

\begin{equation}\label{M2}
C^{-1}_{N\beta}=({2\pi\over \beta})^{N/2}\, \beta^{-{\beta N(N-1)/4}}\,\big[\Gamma(1+{\beta\over 2})\bigl]^{-N}
\prod_{j=1}^N\Gamma(1+{\beta\over 2}{j})\,, \ \ \ \ \ \mathrm{at}\  \beta=1,\,2,\,4
\end{equation}

\medskip
\noindent where the constants $C_{N_\beta}$ are normalisation constants while, in the Random Matrix formalism, the values of $\beta=1,2,4$ refer to the so--called \emph{orthogonal}, \emph{unitary} and \emph{symplectic} ensembles respectively. The cases of $YM$ and $QCD$ theories, as we have seen, are relevant to the ``orthogonal value'' $\beta=1$ ${}^{\footnotemark[10]}$ \footnotetext[10]{As can be seen on (\ref{I2}) and (\ref{I1}) though, the value of $\beta=2$ seems to be selected in reference to the definition (\ref{M1}). This is of course due to the peculiar structure of ${\mathrm{Sp}}\,\mathbb{M}$ as displayed in (\ref{A3}).}. One defines,

\begin{equation}\label{M2}
 \biggl[\,\prod_{j=2}^N\int_{-\infty}^{+\infty}\,{\mathrm{d}\Theta_j}\biggr]  \ P_{N_\beta}(\Theta_1,\dots,\Theta_N)\equiv N^{-1}\,\sigma_N(\Theta_1)\,.\end{equation}

\noindent and Wigner's semi-circle law, valid at large $N$ values, is the following statement \cite{Mehta},

\begin{equation}\label{M3}
\sigma_N(\Theta)\longrightarrow \sqrt{2N-\Theta^2}\,, \ \ \mathrm{for}\  -\sqrt{2N}\leq \Theta\leq +\sqrt{2N}\,,\ \  \sigma_N(\Theta)=0\,,\ \mathrm{otherwise\,.}
\end{equation}
 
Using these relations and observing that integrations on the $\Theta_i$s are saturated on the intervals $|\Theta_i|\leq 1$ because of the strong suppression operated by the gaussian terms $\exp-{\beta\over 2}\Theta_i^2$, beyond $\Theta_i=\sqrt{2/\beta}$, upper and lower bounds on $|C_1|$ and $|C_2|$ can be deduced :

\begin{equation}\label{framing}
{2\over \sqrt{N}}\,C_{N,4}^{-1}\leq |C_1|\leq 4\,\cdot\,2^{{N(N-2)/4}}\, C_{N,2}^{-1},\ \ \ \ \ 2\pi\,C_{N,4}^{-1}\leq |C_2|\leq \pi\,2^{{N(N-2)/4}}\,C_{N,2}^{-1}
\end{equation}

\par
A few remarks are in order.
\par
{\textbf{-}} Proceeding in this way, the intractable sum of monomials generated by expanding the Vandermonde determinant of (\ref{M1}) is circumvented within a few lines of calculations.
\par
{\textbf{-}} Moreover, using Wigner's semicircle law appears to be the more appropriate as its universality is being recognised to  extend far beyond the realm of its original derivation \cite{Krajewski}.
\par
{\textbf{-}} Now, whereas (\ref{M3}) is obtained at $N\rightarrow \infty$ while one has $N=32$ in the current case, corrections to the asymptotic limit of $N\rightarrow \infty$ can be calculated in a systematic, algorithmic manner \cite{A8}.
\par

{\textbf{-}} As in the case of $QCD$ \cite {ygtg}, this shows how the powerful Random Matrix formalism connects to the effective locality non-abelian property to allow for well defined and calculable estimates.

\subsubsection{On the way to a numerical outlook}

Relying on the normalisation constants definitions (\ref{M2}) the absolute value of the ratio $C_1/C_2$ can eventually be bound as follows,

\begin{equation}\label{x}
{2^{1+N-\frac{5}{8}N^2}\over \pi\sqrt{N}}\,(\frac{2}{\sqrt{\pi}})^{N/2}\,\prod_{j=1}^{N\over 2}\Gamma(j+{1\over 2})\leq |{C_1\over C_2}|\leq {2^{1-\frac{N}{2}+\frac{3}{8}N^2}\over \pi}\,(\frac{\sqrt{\pi}}{2})^{N/2}\,\prod_{j=1}^{N\over 2}{1\over \Gamma(j+{1\over 2})}
\end{equation}

Numerically, the numbers involved in both sides of (\ref{x}) are still enormous numbers. That is, besides an issue of existence which is definitely established, (\ref{x}) wouldn't be very helpful for a comparison to numerical estimates.
\par
Rather, taking again advantage of the strong suppression operated by the gaussian terms $\exp-{\beta\over 2}\Theta_i^2$ beyond $\Theta_i$ values greater than $\Theta_i=\sqrt{2/\beta}$, a much easier approximation consists to ignore the squared powers of the variables $(\Theta_k^2- \Theta_l^2)^2$ in the expressions (\ref{I2}) and (\ref{I1}) corresponding to $C_2$ and $C_1$, replacing them by $(\Theta_k- \Theta_l)^2$. In this way, using Wigner's semi-circle law (\ref{M3}), one obtains simply,

\begin{equation}\label{simple}
|\frac{C_1}{C_2}|\simeq\frac{1}{\pi}\,\sqrt{\frac{2}{N}}
\end{equation}

\noindent and one may expect to bring (\ref{205}) in a form more amenable to possible comparisons with lattice numerical simulations, as will be seen shortly (Subsection D),

\begin{equation}\label{2050}
\langle T(A_{\mu}^a(x)A_{\nu}^b(y))\rangle \simeq\frac{1}{2\sqrt{i}}\,\sqrt{{N\over 2g^2N_c}}\,   \sqrt{{\Delta}}\, {\delta^{(4)}(x-y)}\ \delta^{ab}\,{g_{\mu\nu}}\ +\,{\mathcal{O}}({1\over g^2})
\end{equation}

\noindent where the sign of approximate equality refers to (\ref{simple}) and to the use of the Wigner semi-circle law (\ref{M3}).

\subsubsection{Another (drastic) way of estimating the gluon propagator}

Since the first article on the $EL$ matter \cite{EPJC}, much more approximated and simplified technics have been used for computing amplitudes with surprising good results~\cite{5}. The trick consists in substituting a single positive variable, $R$, say, for the involved structure of $(f\cdot\chi)^{ab}_{\mu\nu}(x)$, averaging over all colour and spacetime indices. A crude approximation which may find justifications in some cases, as the one of elastic quark--quark scattering amplitudes for instance. Concerning the $2$--point gluon Green's function, it works : the measure $\mathrm{d}[\chi]$ goes over into $R^7\mathrm{d}R$ and $\det(f\cdot\chi)$ into $R^8$.

\par
Within this approximation the first order gluon $2$--point Green's function is found to be

\begin{equation}
\langle T(A_{\mu}^a(x)A_{\nu}^b(y))\rangle={\mathcal{N}}_0\,\frac{i}{g}\delta^{(4)}(x-y)\,\delta^{ab}g_{\mu\nu}\int_0^\infty \frac{R^7\mathrm{d}R}{R^4}\,e^{\,\frac{i}{4}\Delta R^2}\,(\frac{1}{R})
\end{equation}

with the same meshing parameter $\Delta$ as introduced in (\ref{201}) and (\ref{2011}), and a normalisation factor which reads simply

\begin{equation}
{\mathcal{N}}_0=\int_0^\infty R^3\mathrm{d}R\,e^{\,\frac{i}{4}\Delta R^2}
\end{equation}

Both integrals are straightforward and yield

\begin{equation}\label{Pidi}
\langle T(A_{\mu}^a(x)A_{\nu}^b(y))\rangle=\frac{1}{g}\,\frac{\sqrt{i\pi}}{4}\,\sqrt{\Delta}\, \delta^{(4)}(x-y)\,\delta^{ab}g_{\mu\nu}
\end{equation}

\noindent where the $\delta^{ab}g_{\mu\nu}$ structure comes from the full treatment leading to our main result (\ref{205}), while the dimensional factors of $\sqrt{\Delta}\, \delta^{(4)}(x-y)$ are correctly reproduced. 
\par
A difference of phase remains with a factor of $\sqrt{i}$ instead of its inverse in (\ref{205}). Its origin is easily traced back to this approximation's \emph{artefact}: In effect, replacing ${\left({\mathbb{M}}^{-1}(x)\right)}^{ab}_{\mu\nu}$ by $1/R$, misses the factor of $-i$ which comes from the integration on $O_N(\mathbb{R})$ of the right--most piece of (\ref{201}) to yield (\ref{ON}). Of course such an approximation cannot always be a reliable one: in the case of the current paper, for example, a conclusion like that of Equation (\ref{Amu0}) was far from being reached in this way.

\subsubsection{Proposing a physical picture and a numerical estimate of the meshing parameter}

The main result (\ref{205}) is correct at the dimensional level but its interpretation is not obvious. This is due to the product of the non zero meshing parameter $\sqrt{\Delta}$ with the constraint of  $\delta^{(4)}(x-y)$. These 2 elements seem to be contradictory.
\par
In effect, if $y$ must be \emph{as close as possible} to $x$, then $EL$ has nothing to bring about, but Perturbation Theory only. At face value, the association in a product of the $EL$--length scale, with a constraint of  $\delta^{(4)}(x-y)$ appears as a quirk. 
\par
We will venture here that it isn't so. 
\par Things may be understood as follows. The $YM$ quantum system is originally postulated on the spacetime manifold $\mathcal{M}$ where are labelled the 2 points $x$ and $y$. As the elementary gluonic excitations are integrated out, one leaves the perturbative for the non--perturbative regime of the $YM$ theory, whose interactions are now mediated by the $\chi^a_{\mu\nu}$--fields, whereas references to the same spacetime manifold $\mathcal{M}$ keeps applying unmodified.
\par
This is where the contradiction comes from. In the non--perturbative $YM$ regime, such as described by any \emph{effective form} ${}^{\footnotemark[11]}$, \footnotetext[11]{Not the $EL$ form only, but any other effective form proceeding from functional short distance scales integrations.} $y$ can no longer be taken \emph{as close as possible} to $x$, but a certain minimal distance away from it, here related to $\Delta$.
\par
One has learned to know that functional integrations ``do the job, but do not warn that they do it" ${}^{\footnotemark[12]}$. \footnotetext[12]{Two well known examples are that of the \emph{time ordering prescription} and the \emph{chiral anomaly}, which (quite unexpectedly at first) are both automatically taken care of through the functional integration process.} 
\par\medskip
In the current situation the functional integration process delivers no explicit information
that in the effective formulation (\ref{ZYM40}) the same original points $x$ and $y$ should be understood as \emph{rescaled} points $X$ and $Y$, fuzzed from the original $x$ and $y$ points to an extent $\Delta$.
\par
 That is, $\delta^{(4)}(x-y)$ should be understood as $\delta^{(4)}(X-Y)$ and the $EL$ form (\ref{ZYM40}) postulated on a new spacetime manifold, made out of such points $X,Y,Z,\dots$.
 \par
 This interpretation is reminiscent of the \emph{resolution power} of the microscopes being used to describe the $YM$ dynamics; or, in a more formal way, reminiscent of the Wilson renormalisation group and related block--spin integration procedures used for Ising models \cite{mlb}.
 \par
 A consequence of this interpretation of the result (\ref{205}) amounts to re-write it as

 \begin{equation}\label{2051}
\langle T(A_{\mu}^a(x)A_{\nu}^b(y))\rangle ={\pi N\over 4\sqrt{g^2N_c}}\ \frac{C_1}{C_2}\ \delta^{ab}\,{g_{\mu\nu}}\  \frac{1}{\sqrt{\Delta}}\,{\delta^{(4)}(\widehat{X}-\widehat{Y})}\ \ +\,{\mathcal{O}}({1\over g^2})
\end{equation}

with $X\equiv\Delta\widehat{X}$ so that $\delta^{(4)}(\widehat{X}-\widehat{Y})$ is dimensionless. 
\par
In this way, the gluonic condensate which can be read off (\ref{205}) is seen to go like the inverse squared root of the $EL$--resolution or meshing parameter $\Delta$; that is like $\Lambda^2_{YM}$ at most, if one keeps sticking to an estimation of $\Delta\Lambda_{YM}^4\geq 1$.
\par\medskip
In the literature, not so many numerical estimates can be found based on the gluon propagator and relating to a possible value of the gluon condensate ${}^{\footnotemark[13]}$. \footnotetext[13]{Calculations of gluon condensates are rather based on $\langle({F_a^{\mu\nu}}_R)^2\rangle$ which is gauge invariant from the onset.} In \cite{Kondo}, though, a numerical simulation suggests a value of $(2.76\, GeV)^2$ for the expression $g_R^2\langle \mathcal{A}_\mu^2\rangle_R$ in the case of $SU_c(3)$ \cite{26}, where the subscript $R$ refers to ``renormalised'' (which are all of the expressions dealt with in any $EL$ calculation \cite{QCD5}). 
\par
Now, up to a numerical factor of $N_c/N$ and based on an \emph{Operator Product Expansion} calculation it turns out that in the \emph{Landau gauge}, the term $g_R^2\langle \mathcal{A}_\mu^2\rangle_R$ is the mass acquired by all gluons and \emph{ghost fields} due to gluon condensation; while, remarkably enough, these masses preserve both BRST symmetry and Slavnov-Taylor identities \cite{Kondo}. This opens the route to a sensible comparison of our main result (\ref{205}) and/or  (\ref{2051}) to the numerical simulation just mentioned.
\par\medskip
To begin with, if in (\ref{Pidi}) the strong coupling constant $g_R$ is taken at a value of $10$, then it is quite amusing to discover that the inverse squared root of the meshing parameter $\Delta$ comes out to be

\begin{equation}\label{Pidii}
\frac{1}{\sqrt{\Delta}}=(230\,MeV)^2
\end{equation}

that is, close enough to the $YM$ scale of $\Lambda_{YM}=259\,MeV$ found by a full lattice $QCD$ calculation at zero flavor number $n_f=0$, \cite{Lattice}.
\par
 Using our main result (\ref{205}) now, approximated by (\ref{2050}), the same value of the inverse squared root of the meshing parameter (\ref{Pidii}) is obtained at a smaller value of the strong renormalised coupling constant, that is at $g_R\simeq 4$ which is still, admittedly, a strong coupling value.
\par
As advocated in \cite{Kondo}, a result like (\ref{205}), which can be phrased as ``gluon condensation'', is proposed to be the origin of \emph{mass gap} in the $YM$ theory and of \emph{quark confinement} in $QCD$,

 \section{Conclusion}

In this paper the property of effective locality is explored in the case of pure Minkowskian Yang--Mills theory. As compared to $QCD$ where the property was first discovered, the Yang--Mills case is simpler and preserves some of the most salient features of the non--abelian dynamics, perturbative and non--perturbative. 
\par
In a recent article \cite{miely} the Green's function generating functional of the Yang--Mills theory has been constructed in full details, relying on standard proven functional methods, the Equation (\ref{ZYM40}) of the current paper, whose derivation has been quickly summarised.
\par\medskip
The very purpose of this article is to see wether the formal derivation of $Z_{YM}[j]$ in Equation (\ref{ZYM40}), lends itself to doable concrete calculations. This calls for quite technical developments which, if not the simplest, can nevertheless be carried out with a satisfying degree of mathematical rigour. In particular, as in the case of $QCD$, it turns out to be possible to rely on the powerful \emph{Random Matrix Theory} in order to complete an integration process and to circumvent the untractable difficulties induced by a complicated {\textit{Vandermonde determinant}}.
\par\medskip
The first and elementary gluon Green's function calculation is that of $\langle A^a_\mu(x)\rangle$, the vacuum expectation value of the gluon field. The proof that it is vanishing, {\textit{i.e.}}, $\langle A^a_\mu(x)\rangle=0$, is not trivial, but allows one to set up the general framework of subsequent Green's function calculations.
\par\medskip
This framework is then used to investigate the gluon propagator in the non--perturbative strong coupling regime of the Yang--Mills theory. At the leading order of a strong coupling expansion (in powers of the inverse coupling $1/g$), it is found that $\langle T\,A^a_\mu(x)A^b_\nu(y)\rangle$ comes out proportional to a factor of $\delta^{(4)}(x-y)$ and hence, displays no propagation, contrarily to the perturbative regime. Since the result exhibits also an \emph{isotropy} in the internal and spacetime indices, the non--perturbative gluon propagator could point out the formation of a gluonic condensate of dimension ``mass--squared''. 
\par
It is interesting to observe that a similar conclusion is met within a thermodynamical analysis of quantum $YM$ theory \cite{Ralf}; namely, an exclusion of gluon propagation and the emergence of a massive structure (there identified to emergent fermionic--selfintersecting center--vortex loops), strolling above a confining ground state. Since, in all possible situations, the $T\rightarrow 0$ temperature limit has always been found to be a continuous one, this similarity is worth quoting.
\par
As seen from Perturbation Theory now (renormalisation group--improved), results (\ref{205}) or (\ref{2051}) compare reasonably well with the non vanishing mass acquired by ghosts and gluons in the Landau gauge of the Lorentz--type gauge fixing condition \cite{Kondo, 26}. 
\par
It is stressed that $YM$ effective locality calculations cannot make sense in the absence of a mass or distance scale which turns out to be given by the meshing parameter $\Delta$ introduced in a canonical way in order to define the functional measure of integration on the $\chi$--space of functions defined on the spacetime manifold $\mathcal{M}$. One finds in effect a result of form $\langle T\,A^a_\mu(x)A^b_\nu(y)\rangle\sim \sqrt{\Delta}\,\delta^{(4)}(x-y)$. Would the parameter $\Delta$ be taken to zero, the supposed gluon condensate would vanish altogether, in agreement with the short distance perturbative regime of the $YM$ theory. 
\par
That is, like in $QCD$ \cite{ygtg}, the property of effective locality is clearly associated to a given mass scale beyond which the $EL$ property just fades away, leaving the place to the perturbative regimes of both theories, $YM$ and $QCD$. 
\par
Because of the length scale dependence of the $EL$ property, one may think that contrarily to appearances (such as advocated decades ago in the euclidean case \cite{RefF}), the effective locality form displayed in Equation (\ref{ZYM40}) should hardly be thought of as a form really \emph{dual} to the original $YM$ formulation.
\par\medskip
Now, this paper offers another example of effective locality calculations whose connections to the Random Matrix calculus, by means of analytical continuation, allow one to proceed in a \emph{systematic} way and a satisfying level of  mathematical rigour. It could be worth examining if sub--leading order contributions, $\mathcal{O}(1/g^2)$, would display residual propagation effects, possibly associated to some \emph{glueball} picture. 
\par 
Be it as it may, this treatment of the pure glue sector of $QCD$ should allow one to improve on previous $QCD$ estimates \cite{ygtg} where, to begin with simpler calculations, these gluonic terms had been omitted to the exclusive benefit of fermionic ones.

\par


\textbf{References}
\begin{thebibliography}{**}

\bibitem{EPJC}
  Fried, H.M.; Gabellini, Y.; Grandou, T.; Sheu, Y.--M. Gauge Invariant Summation of All QCD Virtual Gluon Exchanges. \textit {Eur. Phys. J. C} \textbf{2010}, \textit{65}, 395.
 
 \bibitem{tg} 
Fried, H.M.; Grandou, T.; Hofmann, R. On the non--perturbative realization of QCD gauge--invariance. \textit{Mod. Phys. Lett. A} \textbf{2017}, \textit{32}, 1730030.

\bibitem{SW} 
Seiberg, N.; Witten, E. Monopole Condensation, And Conﬁnement In N = 2 Supersymmetric Yang--Mills Theory. \textit{Nucl. Phys. B} \textbf{1994}, \textit{426}, 19; Erratum: \textit{Nucl. Phys. B} \textbf{1994}, \textit{430}, 485. 

\bibitem{RefF}
Reinhardt, H.; Langfeld, K.; von Smekal, L. Instantons in field strength formulated Yang--Mills theories. \textit{Phys. Lett. B} \textbf{1993}, \textit{300}, 111. 

\bibitem{casimir}  
Fried, H.M.; Grandou, T.; Hofmann, R. Casimir operator dependences of non--perturbative fermionic QCD amplitudes.\textit{ Int. J. Mod. Phys. A} \textbf{2016}, \textit{31}, 1650120.

\bibitem{Dmitrasinovic} 
Dmitrasinovic, V. Cubic Casimir operator of $SU_c(3)$ and confinement in the nonrelativistic quark model. \textit{ Phys. Lett. B} \textbf{2001}, \textit{499}, 135.

\bibitem{ygtg}
Fried, H.M.; Gabellini, Y.; Grandou, T.; Tsang, P.H. QCD Effective Locality: A Theoretical and Phenomenological Review. \textit{Universe} \textbf{2021}, \textit{7}, 481. 

\bibitem{miely} 
Fried, H.M.; Gabellini, Y.; Grandou, T. Effective Locality in the pure gluon sector.  \textit{Mod. Phys. Lett. A}  \textbf{2023}, \textit{38}, 2350163.

\bibitem{Eik} 
Fried, H.M. \textit{Basics of Functional Methods and Eikonal Models}; Editions Fronti\`eres: Paris, France, 1990.

\bibitem{Kondo} 
Kondo,  K.-I. Vacuum condensate of mass dimension 2
as the origin of mass gap and quark confinement. \textit{ Phys. Lett. B} \textbf{2001}, \textit{514}, 335.

\bibitem{Halpern}
Halpern, M.B. Field--strength formulation of quantum chromodynamics. \textit{Phys. Rev. D} \textbf{1977}, \textit{16}, 1798.

\bibitem{HS} 
Stratonovich, R.L. On a Method of Calculating Quantum Distribution Functions. \textit{Soviet Physics Doklady} \textbf{1958}, \textit{2}, 416. Hubbard, J. Calculation of Partition Functions. \textit{ Phys. Rev. Lett.} \textbf{1959}, \textit{3}, 77.

\bibitem{Gribov} 
Gribov, V.N. Quantization of non--Abelian gauge theories. \textit{ Nucl. Phys. B} \textbf{1978}, \textit{139}, 1.

\bibitem{QCD6} 
Fried, H.M.; Grandou, T.; Sheu, Y.-M. Non--Perturbative QCD Amplitudes at Quenched and Eikonal Approximations. \textit{Ann. Phys.}  \textbf{2014}, \textit{344}, 78, Appendix A.

\bibitem{Mehta} 
Mehta, M.L. \textit{Random Matrices}; Academic Press: Cambridge, MA, USA, 1967.
\bibitem{Ted}
G.W. Johnson and M.L. Lapidus, \textit{The Feynman Integral and Feynman's Operational Calculus}, Oxford University Press (2000).

\bibitem{Rodrigues} 
Friedberg, S.; Insel, A.; Spence, L. \textit{Linear Algebra};  Pearson Education, Inc:  Upper Saddle River, NJ, USA, 2003.

\bibitem{Zhang} 
Zhang, L.  Volumes of Orthogonal Groups and Unitary Groups. \textit{arXiv} \textbf{2017}, arXiv:math-ph/1509.00537v5.

\bibitem{Douglas} 
Douglas, M. Report on the Status of the Yang--Mills Millennium Prize Problem \textbf{2004}, https://ncatlab.org/nlab/files/Douglas-StatusOfYMMillenniumProb.pdf

\bibitem{Zee}
 Zee, A. \textit{Quantum Field Theory in a Nutshell};  Princeton University Press: Princeton, NJ, USA, 2010; chap.I.3.

\bibitem{tgpt}
 Grandou, T.; Tsang, P.H.  Effective Locality and Chiral Symmetry Breaking in QCD. \textit{Mod. Phys. Lett. A} \textbf{2019}, \textit{34}, 1950335.

\bibitem{Krajewski}
 Krajewski, T.; Tanasa, A.; Vu, D.L. Wigner law for matrices with dependent entries--a perturbative approach. \textit{J. Phys. A: Math. Theor} \textbf{2017}, \textit{50}, 16LT02.

\bibitem{A8} 
See Ref.\cite{Mehta}, Appendix \textbf{A8}.

\bibitem{5} 
Fried, H.M.; Tsang, P.H.; Gabellini, Y.;  Grandou, T.;  Sheu, Y.-M. Comparison of QCD curves with elastic pp scattering data. \textit{Int. J. Mod. Phys. A} \textbf{2019}, \textit{34}, 1950236.

\bibitem{mlb} 
Le Bellac, M. \textit{Des ph\'enom\`enes critiques aux champs de jauge};  InterEditions/Editions du CNRS: Paris, France, 1988; p. 91.

\bibitem{26} 
Boucaud, Ph.; Burgio, G.; Di Renzo, F.; Leroy, J.P.; Micheli, J,; Parrinello, C.; P\`ene, O.; Pittori, C.; Rodriguez-Quintero, J.; Roiesnel, C.; Sharkey, K. Lattice calculation of $1/p^2$ corrections to $\alpha_s$ and of $\Lambda_{QCD}$ in the MOM scheme. \textit{JHEP 0004: 006} \textbf{2000}.

\bibitem{QCD5}
Fried, H.M.; Tsang, P.H.; Gabellini, Y.;  Grandou, T.;  Sheu, Y.-M. An exact, finite, gauge--invariant, non--perturbative approach to QCD renormalization. \textit{Ann. Phys.}
\textbf{2015}, \textit{359} 1.

\bibitem{Lattice}
G\"ockeler, M.; Horsley, R.; Iving, A.C.; Pleiter, D.; Rakow, P.E.L.; Schierholz, G.; St\"uben, H. Determination of the Lambda parameter from full lattice QCD. \textit{Phys. Rev. D} \textbf{2006}, \textit{73}, 014513.

\bibitem{Ralf}
 Hofmann, R. \textit{The Thermodynamics of Quantum Yang Mills Theory};  World Scientific: Singapore, 2012.



\end{thebibliography}
\end{document}